\documentclass[notitlepage,prd,nofootinbib,longbibliography,a4paper,11pt,eqsecnum]{revtex4-1}
\usepackage{amsmath} 
\usepackage{graphicx} 
\usepackage{amsthm}
\usepackage{amssymb} 
\usepackage{dsfont}
\usepackage{yfonts}
\usepackage{hyperref}
\usepackage{array,xcolor,graphicx}
\usepackage{booktabs,multirow}
\usepackage[utf8]{inputenc}
\usepackage{mathtools}

\usepackage{slashed}

\usepackage{etoolbox}
\patchcmd{\section}
  {\centering}
  {\raggedright}
  {}
  {}
\patchcmd{\subsection}
  {\centering}
  {\raggedright}
  {}
  {}
\usepackage[all]{xy}
\usepackage{tikz}
\usetikzlibrary{arrows.meta}

\hypersetup{colorlinks=true,linkcolor=blue,citecolor=red,urlcolor=blue}

\newcommand{\be}{\begin{equation}}
\newcommand{\ee}{\end{equation}}
\newcommand{\bea}{\begin{eqnarray}}
\newcommand{\eea}{\end{eqnarray}}

\def\tr{{\mathrm{tr}}}

\def\d{\partial}

\def\CB{\mathcal{B}}

\def\CM{\mathcal{M}}
\def\CN{\mathcal{N}}

\def\CT{\mathcal{T}}
\def\CO{\mathcal{O}}
\def\CP{\mathcal{P}}

\def\muR{\boldsymbol{\mu}_R}

\begin{document}
    
\title{Relativistic hydrodynamics with the parity anomaly}

\author{Napat Poovuttikul }
\affiliation{Department of Mathematical Sciences, Durham University, Durham, DH1 3LE, United Kingdom}
\begin{abstract}
\vspace{1cm}
We consider the hydrodynamic regime of a 2+1 dimensions QFT with the parity anomaly. Beyond the known constraints from positivity of entropy production, we show that the anomaly inflow mechanism, from a corresponding bulk SPT phase, together with thermodynamic consistency of equilibrium partition functions, restricts the form of non-dissipative transport coefficients. This included the known form of quantised Hall conductivity, which is fixed to be $\sigma_{xy} = e^2/2h$, along with new constraints on other three non-dissipative parity-odd transport coefficients. 
\end{abstract}
\maketitle
    \begingroup
    \hypersetup{linkcolor=black}
\newpage
    \tableofcontents
    \endgroup
\section{Introduction and motivations}

The intertwining of hydrodynamic and quantum regimes of quantum field theory provides an interesting playground of interesting mathematical ideas with observable consequences. This regime can be found in both (very clean) condensed matter system \cite{Bandurin2016,Crossno1058,moll2016evidence} and in quark-gluon plasma \cite{Romatschke:2007mq} (see also \cite{Romatschke:2017ejr} for recent extensive review). Symmetry structures of quantum systems thereby enrich the ordinary hydrodynamics to be much more interesting than, arguably, conservation of energy-mentum with extra steps. 

Perhaps one of the most interesting developments in this direction is the interplay between quantum anomaly and hydrodynamics. 
Take a hydrodynamic limit of a QFT with $U(1)^3$ 't Hooft anomaly\footnote{For standard reviews on anomaly, see e.g. \cite{Harvey:2005it,Bilal:2008qx} or \textsection 13 of \cite{nakahara2018geometry}} in $3+1$ dimensions whose Ward identity, in the presence of background gauge field $a_\mu$, can be written as 
\begin{equation}\label{eq:U(1)3AnomWardIden}
  \d_\mu j^\mu = \kappa \epsilon^{\mu \nu \rho \sigma}(da)_{\mu \nu}(da)_{\rho \sigma}\,.
\end{equation}
This anomaly will remain fixed all the way to the low energy description of this theory \cite{tHooft:1979rat}. 
According the hydrodynamic principle (see e.g. \cite{LLfluid,forster1995,Kovtun:2012rj}), one may attempt to write down a parity-odd (due to anomaly) current $j^\mu$ in terms of hydrodynamic variables -- the temperature $T$, the chemical potential $\mu$ and fluid velocity $u^\mu$ order by order in the gradient expansions. The most general terms one can write down, up to the first derivative level is 
\begin{equation}\label{eq:Son-surowka-constiutive}
  j^\mu = \rho u^\mu + \xi \,\epsilon^{\mu \nu \rho \sigma} u_\nu \nabla_\rho u_\sigma + \frac{1}{2}\xi_B \,\epsilon^{\mu \nu \rho \sigma} u_\nu (da)_{\rho \sigma}+...
\end{equation}
where we omitted the usual dissipative first derivative terms (interested readers can find the explicit relations in the above textbook references). \textit{Intriguingly, the parity odd transport coefficients $\{\xi,\xi_B \}$ are completely fixed in terms of anomaly coefficients $\kappa$ and thermodynamic variables}. This result can be derived from purely macroscopic arguments such as positivity of entropy production \cite{Son2009}, properties of thermal correlation function \cite{Jensen:2012jy}, consistency of equilibrium partition function \cite{Banerjee:2012iz} and hydrodynamic effective action \cite{Dubovsky:2011sk,Haehl2014,Glorioso:2017lcn}. This has a vast impact on the studies of transport phenomena that we cannot do justice here\footnote{We recommend the reader to a recent review \cite{Landsteiner:2016led} for further informations and key references on this aspect.}.
Many generalisation of the anomaly constraints on macroscopic transport have been made over the years such as generalisation of $U(1)$ to arbitrary group Lie group $G$, inclusion of gravitational anomaly and different kind of 't Hooft anomaly across arbitrary (even)dimensions, see e.g. \cite{Neiman:2010zi,Banerjee:2012cr,Jensen:2013kka}.

The origin of all anomalies of continuous group $G$, types discussed above, can be traced back to existence of chirality that is exclusive to even spacetime dimensions (such as chiral fermion in the seminal ABJ anomaly). These are not only anomalies in nature. By unrestricted ourselves from continuous group, one finds that there can be anomaly in both odd and even spacetime dimensions and the matter may or may not involves massless fermion \cite{Wen:2013oza,Kapustin:2014lwa,Kapustin:2014zva}\footnote{
  We also find an entertaining and insightful \href{https://youtu.be/QiMxPE3wUsE}{\texttt{lecture series}} by Y. Tachikawa to be a good start for familiarising the concept of discrete anomaly.}. The oldest and most well-known (albeit a lot less than its continuous symmetry counter part) is the parity anomaly in $2+1$ dimensions \cite{Redlich:1983kn,Niemi:1983rq,AlvarezGaume:1984nf}. Some of its notable properties, in comparison to the usual anomalies, should be mentioned here 
\begin{itemize}
  \item While anomalies of continuous group can be detected through perturbative computation (such as triangle diagram in $3+1$ dimensions), the information of parity anomaly and the kind involving discrete anomaly cannot be obtained in such a way. Hence there is a nomenclature, perturbative and non-perturbative anomaly for the former and latter case respectively.
  \item In the langauge of partition function, one can define anomaly by how it transformed under continuous background gauge transformation e.g. 
\begin{equation}
  Z[a+d \lambda] = Z[a] \exp\left( i\Phi[\lambda, da]\right)
\end{equation}
  where $\Phi[\lambda, da]$ is a phase ambiguity of the partition function and $\lambda$ is the transformation parameter of the background gauge field. In contrast, the partition function of a theory with parity anomaly in such a way that is invariant under $a\to a + d \lambda$. However, it does transformed under time-reversal $Z\to Z^\CT$, namely
\begin{equation}
  Z^\CT = Z \exp(-i\pi\boldsymbol{\eta})
\end{equation}
  where the phase ambiguity is the Atiyah-Patodi-Singer (APS) $\boldsymbol{\eta}-$invariant \cite{atiyah1975spectral}\footnote{In a simple term, it is a (regulated) difference between positive and negative eigenvalues of the Dirac operator $i\slashed{D}$ on a given manifold.}. As a result, this anomaly does not show up in the anomalous Ward identity, unlike the example in \eqref{eq:U(1)3AnomWardIden} and its generalisation.

  \item The partition function of anomalous theory can be made invariant when treated as a boundary of a topological field theory in one higher dimensions (through anomaly inflow mechanism \cite{Callan1985}). For continuous symmetry in $d+1 \in 2\mathbb{Z}$ dimensions, the bulk is described by Chern-Simons term in $d+2$ dimensions. There is no Chern-Simons term in even $d+2$ dimensions, however. As it turns out, the bulk theory that cancel the parity anomaly of the boundary QFT is\footnote{\label{fn:conventionF}We use the convention where $\oint F = 2\pi \mathbb{Z}/e$. The prefactor $e^2/32\pi^2$ is chosen such that, on a closed manifold $S_{top} = \theta n$ with $n\in \mathbb{Z}$.
  This kind of bulk theory is often referred to as SPT phase. Unlike Chern-Simons theory, it has a unique ground state. Its partition function can be written as a complex number $Z_{bulk} = \exp(i\Phi)$ of dimension 1 and its inverse is simply a complex conjugate $Z_{bulk}^{-1} = \exp(-i\Phi)$, see \cite{Witten:2015aba} for similar discussion.
  } 
\begin{equation}\label{eq:bulkSPT}
  S_{bulk} = \frac{\theta e^2}{32\pi^2\hbar }\int d^4X \, \epsilon^{abcd} F_{ab}F_{cd}\qquad \text{at}\qquad \theta = \pm \pi
\end{equation}
where $F$ is the field strength of the (boundary) background gauge field $a$ extended to the bulk. As it turns out, this bulk action is an effective description of a topological insulator \cite{Qi:2008ew}. Note that, for a fixed $\theta$, this bulk action \eqref{eq:bulkSPT} is odd under time-reversal but the combined bulk + boundary system is invariant through the discrete version of inflow mechanism. In this particular case, the cancellation is a manifestation of the APS index theorem. The reader may find a (very)brief demonstration of how APS index works in this particular case in Appendix \ref{app:APSindex}.
\end{itemize}
All these information, particularly the part concerning anomaly inflow were nicely explained in \cite{Witten:2015aba,Witten:2019bou}. We highly recommend to readers for more detailed explanation. Studies on effects of perturbative anomaly in the effective theory description in a setup closely related to ours can be found in \cite{Chan:2015zhy,Golkar:2015oxw}.

The goal this work is to understand whether or not the parity anomaly can fix, \textit{without resorting to microscopic details}, the value of parity-odd hydrodynamics in $2+1$ d in the same manner as its perturbative anomaly counter parts in even dimensions.
It should be note that, since the parity anomaly does not altered the Ward identity, such constraint cannot be derived using positivity of entropy production argument of \cite{Son2009}. To see the effect of parity anomaly, we restrict ourselves to the hydrodynamic equilibrium partition function approach of \cite{Jensen2012}. The technology to study anomaly inflow in such configuration was developed in \cite{Jensen:2013kka} and we simply apply it to the current context of parity anomaly. With this method, we show that there are constraints on parity-odd fluid transport coefficients as a consequence of thermal equilibrium and anomaly inflow. There is no microscopic details required other than that parity anomaly persist all the way from the UV to the hydrodynamic regime in the deep IR.

Next in Section \ref{sec:result}, we outlined what is the parity-odd fluid in 2+1 dimensions (its transport coefficients and previously known constraints) then stating our result on which of these transport coefficients are fixed by parity anomaly. One of the constraint resulting in the fixed value of Hall conductivity 
\begin{equation}\label{eq:HallConduct}
  \sigma_{xy} = \frac{e^2}{2h}\,.
\end{equation}
where the factor $e^2$ appears as a result of quantisation condition of flux $\oint F = 2\pi \text{ (integers)}/e$ in \eqref{eq:bulkSPT}.
This is a well-known result obtained form the free theory computation at the boundary of topological insulator (see e.g. Section III of \cite{Qi:2011zya} and Section V.A. of \cite{Hasan:2010xy}). Others transport coefficients that are fixed by anomaly will be elaborate in the same section.
 The derivation of these results are provided in Section \ref{sec:derivation}. We will discuss the result, possible applications and future directions in Section \ref{sec:discussion}.

\section{Parity-odd fluid, thermal equilibrium and results}\label{sec:result}

By fluid or hydrodynamics, we do mean the conventional definition classic textbooks such as \cite{LLfluid,forster1995}. It is an IR effective description a gapless theory at finite temperature and possibly chemical potential/density. A possible justification for hydrodynamic limit in a QFT is that, at finite temperature, generic operators that decays away at late-time and long-distance, except those that are commuted with the Hamiltonian.   
In a relativistic system with Poincar\'e and $U(1)$ global symmetry, it means that all conserved currents $T^{\mu \nu},j^\mu$ can be written in terms of $\{T, u^\mu, \mu\}$ which are conjugate to energy, momentum and $U(1)$ charge\footnote{
This should be contrasted with other effective theory, also dubbed hydrodynamics, but do not include temperature, fluid velocity as well as dissipative effects in the description, such as in \cite{Chan:2015zhy}. The equilibrium sector of such EFT can be obtained from what described here upon turning off the temperature and fixing the fluid velocity at rest. 
}. Expressions for $T^{\mu \nu},j^\mu$ in terms of these variables are called constitutive relations. For a parity-invariant fluid, the terms up to first order in the derivative expansions and their constraint from positivity of entropy production have been classified and can be found in \cite{LLfluid}. The profiles of these hydrodynamic variables can be obtained by solving the Ward identities 
\begin{equation}
  \nabla_\mu T^{\mu \nu} = (da)^{\nu \mu}j_\mu \, , \qquad \nabla_\mu j^\mu = 0\,,
\end{equation}
where the theory is defined on a manifold with metric $g_{\mu \nu}$ and background $U(1)$ gauge field $a_\mu$. Below, we briefly review the notations and machinery that used to arrived at the result. 

When parity is broken, by whatever means, the constitutive relations need to be modified to included terms constructed from pseudoscalars, vectors and tensors. It was notice many decades ago that these new terms have consequences in transport phenomena \cite{avron1998odd}. A complete classification of thermodynamically consistent parity odd in 2+1 dimensions was done relatively recently in \cite{Jensen:2011xb,Banerjee:2012iz} (see also \cite{Kaminski:2013gca} for its non-relativistic version and \cite{Lucas:2014sia} for recent discussion on applications). One finds the modified constitutive relations to be\footnote{This constitutive relation chose the form of off-equilibrium fluid variables in such a way that $u_\mu j^\mu = \rho$, $u_\mu u_\nu T^{\mu \nu} = \varepsilon$ and $u_\mu T^{\mu \nu}= (\varepsilon+p)u^\nu$ without any derivative corrections. This choice of variables is known as Landau frame, see \textsection XV of Landau\&Lifshitz's book \cite{LLfluid}. This is not a unique choice of variables and another popular ``frame'' known as the Eckart frame can be found in the rest of the same book.}
\begin{equation}\label{eq:parityOddConstitutiveReln}
  \begin{aligned}
T^{\mu \nu} &= \varepsilon u^\mu u^\nu + \left( p -\zeta \nabla_\lambda u^\lambda 
-\tilde\chi_B B -\tilde\chi_\Omega \Omega  \right) - \eta \sigma^{\mu \nu} - \tilde\eta \sigma^{\mu \nu}\, , \\
j^\mu &= \rho u^\mu + \sigma V^\mu + \tilde \sigma \epsilon^{\mu \nu \lambda} u_\nu V_\lambda + \tilde\chi_E \epsilon^{\mu \nu \lambda}u_\nu E_\lambda + \tilde\chi_T \epsilon^{\mu \nu \lambda}u_\nu \nabla_\lambda T\,.
  \end{aligned}
\end{equation}
Here, the $\varepsilon,p,\rho$ are the usual energy density, pressure and $U(1)$ density whereas $\eta,\zeta$ are the shear and bulk viscosities present in the ordinary fluid. The scalar, vector and tensor structures found above are defined as 
\begin{equation}
  \begin{aligned}
\sigma^{\mu \nu} &= 2\nabla_{(\mu} u_{\nu)} - \Delta^{\mu \nu}\nabla_\lambda u^\lambda\, ,\qquad\quad &
\tilde\sigma^{\mu \nu} &= \frac{1}{2} \left( \epsilon^{\mu \alpha \beta}u_\alpha \sigma_{\beta}^{\;\; \nu} + \epsilon^{\nu \alpha \beta}u_\alpha \sigma_{\beta}^{\;\; \mu} \right)\, , \\
V^\mu &= E^\mu - T \Delta^{\mu \nu}\nabla_\nu\frac{\mu}{T}\, , \qquad\quad &
\tilde V^\mu &=\epsilon^{\mu \alpha \beta}u_\alpha V_\beta\, , \\
E^\mu &= (da)^{\mu \nu}u_\nu\, , \qquad \quad&
\tilde E^\mu &= \epsilon^{\mu \alpha \beta}u_\alpha E_\beta\,,\\
\Omega &=  -\epsilon^{\mu \alpha \beta} u_\mu \omega_{\mu \nu}\,,\qquad \quad &\CB &=  -\frac{1}{2}\epsilon^{\mu \alpha \beta}u_\mu (da)_{\alpha \beta}
  \end{aligned}
\end{equation}
with $\Delta =g^{\mu \nu} +u^\mu u^\nu$ be a projector along the plane perpendicular to $u^\mu$ and $\omega_{\mu \nu} = \Delta^{\mu \alpha}\Delta^{\nu \beta} \nabla_{[\alpha}u_{\beta]}$ is the vorticity. The tilded vectors/tensors are pseudovectors/tensors and $\tilde\eta, \tilde\chi_B,\tilde\chi_\Omega,\tilde\sigma,\tilde\chi_E,\tilde\chi_T$ are additional transport coefficients in the parity-odd fluid. Requiring that the entropy production is positive will implies positivity of terms governing dissipative effect i.e. $\eta \ge 0,\zeta\ge 0$ and $\sigma\ge 0$. There is no constraint on non-dissipative transport coefficients $\tilde\eta,\tilde\sigma$ and but there are relations among $\tilde\chi_B,\tilde\chi_\Omega,\tilde\chi_E,\tilde\chi_T$ \cite{Jensen:2011xb}.

Constraints on the last set of transport coefficient is best understood as a consistency of equilibrium partition function which we will briefly elaborate. It means that there is a time-like Killing vector $K^\mu = u^\mu/T$ and consistent set of sources $g_{\mu \nu}, a_\mu$. This resulting in the vanishing of entropy production, the structures $\sigma^{\mu \nu},\tilde\sigma^{\mu \nu}$ and $V^\mu$, along with the following relations
\begin{equation}\label{eq:hydrostaticIden}
  \nabla_\mu T = -T u^\lambda \nabla_\lambda u_\mu\, , \qquad \d_\mu \mu = -\mu u^\lambda\nabla_\lambda u_\mu+ E_\mu \, , \qquad \nabla_\mu u_\nu = -u_\mu u^\lambda\nabla_\lambda u_\nu + \omega_{\mu \nu}
\end{equation}
Since this configuration describes non-dissipative sector of the theory, it can be captured by the ordinary generating function\footnote{This should be contrast with Closed-Time-Path formalism where the degrees of freedom are doubled in the complex time contour of \cite{Schwinger:1960qe,Keldysh:1964ud} to take into account the dissipative effect and fluctuation-dissipation theorem. See e.g.\cite{kamenev2011field} for a more modern notations. }. There is a nice geometric interpretation to the hydrodynamic variables when analytically continued the direction along $K^\mu$ to the thermal $S^1$ i.e. the temperature and chemical potential is the (proper)size of and the $U(1)$ holonomy around the thermal cycle and $u^\mu$ is the normalised Killing vector along $S^1$ direction. 
 One can then proceed to write down every possible independent (pseudo)scalars out of these geometric quantities up to the first order in the derivative expansions, as done in \cite{Jensen2012,Banerjee:2012iz}, and find the following expression for $W = -i \log Z$
\begin{equation}\label{eq:HydroStaticW}
  W = \int d^3x \left( \sqrt{-g}\,p(T,\mu) + \tilde\alpha_1 \CB  +\tilde\alpha_2 \Omega \right) + \CO(\d^2)\,.
\end{equation}
where $p,\tilde\alpha_1,\tilde\alpha_2$ are, a priori, arbitrary function of $T,\mu$ to be determined by microscopic computations.
It turns out that, all the non-dissipative transport coefficients in \eqref{eq:parityOddConstitutiveReln}, except $\tilde\eta,\tilde\sigma$, are related to $\tilde\alpha_1,\tilde\alpha_2$ in the following way
\begin{equation}\label{eq:WtoLandauFrame}
  \begin{aligned}
\tilde\chi_B &= -\frac{\d p}{\d \varepsilon} \left( T\frac{\d \tilde\alpha_1}{\d T} + \mu \frac{\d \tilde\alpha_1}{\d \mu} - \alpha_1\right)- \frac{\d p}{\d \rho}\frac{\d \tilde\alpha_1}{\d \mu}\, , \\
\tilde\chi_E &= \frac{\d\tilde \alpha_1}{\d \mu} - \frac{\rho}{\varepsilon+p} \left(\frac{\d \tilde \alpha_2}{\d \mu} - \tilde\alpha_1\right)\, , \\
\tilde\chi_\Omega &= - \frac{\d p}{\d \varepsilon} \left( T\frac{\d \tilde \alpha_2}{\d T} + \mu \frac{\d \tilde \alpha_2}{\d \mu} -2 \tilde \alpha_2 \right) - \frac{\d p}{\d \rho} \left( \frac{\d \tilde \alpha_2}{\d \mu}- \tilde \alpha_1 \right)\, , \\
\tilde\chi_T &= \frac{1}{T} \left( T\frac{\d \tilde \alpha_1}{\d T} + \mu \frac{\d \tilde \alpha_1}{\d \mu}- \tilde \alpha_1 \right) - \frac{\rho}{T(\varepsilon+p)} \left( T\frac{\d \tilde \alpha_2}{\d T} + \mu \frac{\d \tilde \alpha_2}{\d \mu} -2 \tilde \alpha_2 \right)\,.
  \end{aligned}
\end{equation}
To the best of our knowledge, these are the most exhausted macroscopic constraints of a consistent parity-odd fluid so far.

Now that we are done with the technicality and terminology, let us state the new result in the work. If the parity-violation can be removed by coupled the fluid to the bulk SPT phase in Eq.\eqref{eq:bulkSPT}, we will show, in the next section, that 
\begin{equation}\label{eq:result-1}
  \tilde \alpha_1 = \mu \left(\frac{\theta e^2}{2\pi h}\right)\, , \qquad \tilde \alpha_2 = \frac{1}{2} \mu^2 \left(\frac{\theta e^2}{2\pi h}\right)\, , \qquad \text{with}\qquad \theta = \pm \pi\,.
\end{equation}
and consequently
\begin{equation}\label{eq:result-2}
  \tilde\chi_\Omega=\tilde\chi_T = 0\, , \qquad \tilde\chi_B = -\frac{\d p}{\d \rho} \theta\, , \qquad \tilde\chi_E = \theta\,.
\end{equation}
in the unit of $e^2/(2\pi h)$. One can immediately see what this implies for a Hall conductivity. We can considering the setup where we have a fixed $u^\mu=(1,0,0)$, which can be obtained in a system with disorder that break translation symmetry, keeping the temperature constant and set $E^i = \d_i \mu$ so that the system remains in equilibrium. As a result, we find that there is an equilibrium Hall current:
\begin{equation}
  j^x = \tilde \chi_E E_y \, , \qquad j^y=-\tilde\chi_E E_x\,.
\end{equation}
Compared with the definition of Hall conductivity $j^i = \sigma_{xy}\epsilon^{ij} E_j$, we find the known value of Hall conductivity on the edge of topological insulator advertised in \eqref{eq:HallConduct}.

\section{Derivation of constraints from parity anomaly}\label{sec:derivation}

By writing the non-dissipative sector of parity-odd fluid in the equilibrium partition function, it is not very surprising that parity anomaly will further restrict the form of coefficients $\tilde\alpha_1$ and $\tilde\alpha_2$ in \eqref{eq:HydroStaticW}. After all, the anomalous can be canceled by anomaly inflow mechanism which is to say that the combined bulk+boundary system is time-reversal invariant
\begin{equation}
  Z_{inv} = Z_{hydro}[a] \exp(iS_{top})\,.
\end{equation}
where $Z_{hydro} = \exp(iW)$ in \eqref{eq:HydroStaticW} and $S_{top}$ is the action for SPT phase in \eqref{eq:bulkSPT}. The challenge is then to write the topological term (in one dimension higher) in terms of hydrodynamic variables. Luckily, such machinery developed in \cite{Jensen:2013kka} for perturbative anomalies can be immediately applied to this case. We will use this approach to derived the results in \eqref{eq:result-1}-\eqref{eq:result-2}. 

Let's consider the topological term \eqref{eq:bulkSPT} on a manifold $\CN$ whose boundary $\d \CN = \CM$ support the parity-odd fluid. Let $u = u_a dX^a$ be a 1-form dual of timelike Killing vector on $\CN$ whose boundary is the fluid velocity $u_\mu dx^\mu$ such that $u_\mu u^\mu = u_a u^a = -1$. Similarly $\mu, \omega = \frac{1}{2} \omega_{ab} dX^a\wedge dX^b, A = A_a dX^a, F = \frac{1}{2} F_{ab} dX^a\wedge dX^b$ are the local chemical potential, vorticity, background gauge field $a$ and field strength $da$ extended from $\CM$ to $\CN$. The bulk is also required to be in equilibrium, that is there is a timelike Killing vector $K^a = u^a/T$ and the chemical potential is the (analytically continued)holonomy $\mu = u^a A_a$ that become $K^\mu = u^\mu/T$ and $\mu =u^\mu a_\mu$ at $\CM = \d \CN$.
One can define another $U(1)$ connection to be 
\begin{equation}
  A^\perp = A + \mu\, u\, ,\qquad \text{such that}\qquad u^a A^\perp_a = 0\,.
\end{equation}
We can define the field strength out of $A^\perp$ via $F^\perp = dA^\perp$. This field strength has vanishing $\int F^\perp\wedge F^\perp$. An easy way to see this is to imagine $u^a$ to be along the time direction and for it to be in equilibrium, we need $A_a$ to be independent of time. Thus $A^\perp$ and $F^\perp$ has no component in time direction and therefore $\epsilon^{abcd} (F^\perp\wedge F^\perp)_{abcd}$ vanished. This means that 
\begin{equation}
  S_{top}\sim \int_\CN F\wedge F = \int \left( F\wedge F - F^\perp\wedge F^\perp \right)
\end{equation}
The r.h.s. is particularly useful since we can invoke the Chern-Weil theorem to which states that the difference between anomaly polynomial corresponds to two different connections (in this case $A$ and $A^\perp$) can be written as an exact form. This means that 
\begin{equation}
  \CP[F]-\CP[F^\perp] = d V_\CP \, ,\qquad \text{where} \qquad \CP[F] = F\wedge F\,,\quad \CP[F^\perp]=F^\perp\wedge F^\perp\,,
\end{equation}
and that $\int_\CN F\wedge F = \int_\CM V_\CP$. The 3-form $V_\CP$ is known as transgression.
More details on this can be found in appendix D of \cite{Jensen:2013kka} or more formal discussion in \textsection 11 of \cite{nakahara2018geometry}\footnote{Had we chose $\CP[F^\perp = 0]$, one will have $\int_\CN F\wedge F = \int_\CM S_{CS}$ which is the usual Chern-Simons form. This is but one representation of the transgression and only a globally well-defined when $F$ can be written as $da$ everywhere at the boundary.   }. 

To determined the transgression $V_\CP$, let consider a parameter $\tau \in [0,1]$ connecting $A$ and $A^\perp$ as 
\begin{equation}
  A(\tau) = A + (1-\tau) \mu\, u
\end{equation}
and that the field strength 
\begin{equation}
  F(\tau) = -u\wedge (E + (\tau-1)d \mu) + (B - 2(1-\tau)\mu\, \omega)\,.
\end{equation}
where $B = F + u\wedge E$ is a 2-form magnetic field. 
Using the commutativity between differential $d$ and derivative along $\tau$ direction $\d_\tau$, one finds that 
\begin{equation}
\begin{aligned}
  V_\CP &= \int^{\tau=1}_{\tau = 0}d\tau \left( \d_\tau A_\tau \wedge \frac{\d \CP[F(\tau)]}{\d F(\tau)} \right)\,,\\
  &= -2 \mu\, u\wedge F - 2\mu^2 \, u \wedge \omega\, ,\qquad\qquad \text{for}\qquad \CP[F(\tau)]=F(\tau)\wedge F(\tau)\,.
\end{aligned}
\end{equation}
Evaluated the transgression at the boundary and convert it to the index notation, we find that 
\begin{equation}\label{eq:StopOnM}
  S_{top} = -\frac{\theta e^2}{4\pi^2 \hbar}\int_\CM d^3x \left( \mu\,  \CB + \frac{1}{2} \mu^2 \,\Omega   \right)
\end{equation}
Combining $S_{top}$ with hydrodynamic equilibrium partition function, we find that, for 
\begin{equation}
-i \log Z_{inv} = W + S_{top}  
\end{equation}
to be invariant, one need cancel the terms odd under time-reversal. That is $\tilde \alpha_1$ and $\tilde \alpha_2$ must follows \eqref{eq:result-1}. Upon obtaining $T^{\mu \nu}, j^\mu$ from $W$ and convert it to the Landau frame, as in \cite{Jensen2012} using \eqref{eq:WtoLandauFrame}, one finds the transport coefficients in \eqref{eq:result-2}. This conclude our derivation.

Readers who familiar with the full form of APS index theorem (or notice the appendix \ref{app:APSindex}) may worry that we have been ignoring the contribution from Euler density. We would like to reassured that it does not contribute to the hydrodynamic transport at this order in the derivative expansions. A verification of this statement can be found in appendix \ref{app:EulerDensity}. 

An important remark also has to be made concerning the periodicity of $\theta-$angle and whether or not presented effect is simply a pure contact term. It can be easily seen that the topological partition function $Z=\exp(iS_{top})$ on $\CN$, if $\CN$ is closed, is invariant under $\theta \to \theta + 2\pi \mathbb{Z}$. Similarly, one may wonder if one simply add a Chern-Simons counter term to $\CM$ and altered or make the transport coefficient vanishes. As it turns out, one can indeed using thermal equilibrium identities \eqref{eq:hydrostaticIden} to write the Chern-Simons action in the form similar to \eqref{eq:StopOnM}. However, the allowed counter term has to be quantised property (see e.g. \cite{Closset:2012vp}) and can only produce a term in \eqref{eq:StopOnM} with $\theta = 2\pi \mathbb{Z}$. One can therefore conclude that the ambiguity of $\theta-$angle from the bulk can be removed by appropriately adjusted counter term and vice-versa. Only the fractionaly part i.e. $\theta/2\pi = \pm 1/2$ is physical and is therefore procedure independent.  

\section{Discussions}\label{sec:discussion}

Let us reiterate new results in the work. Perhaps the most striking features of this analysis is the existence of equilibrium Hall conductivity which is fixed to, $\sigma_{xy} = \tilde\chi_E = \theta e^2/2h$ for $\theta = \pm \pi$, which is a hallmark of topological insulator. Any naive EFT that claims to be in the same universality class as boundary of topological insulator would have to put in this condition, either by hand or certain microscopic inputs. The consistency of anomaly inflow and equilibrium partition function not only obtained this result purely from macroscopic point of view but also put restrictions the other non-disspative transport coefficients. From \eqref{eq:result-2}, we saw that, if the parity anomaly is the only source of broken time-reversal symmetry, then
\begin{itemize}
  \item[(i)] The pressure response to magnetic field $\tilde \chi_B = -(\d p/\d \rho)\theta$ is fixed by thermodynamic functions and $\theta = \pm \pi$.  
  \item[(ii)] The pressure response to vorticity $\tilde\chi_\Omega$ and current responds to perpendicular temperature gradient $\tilde\chi_T$ are forced to vanish.
\end{itemize}
Measurements of these quantities could served as additional indication whether or not a system of interest could have parity anomaly. It should be emphasised here that our analysis only relies on the (anomalous)global symmetries in the IR, thermodynamic consistencies and assumption of gradient expansions. Thus, we expect this result to be robust even when the microscopic descriptions involved strong interactions.

Another way the system can be realised without being at the boundary of $3+1$ bulk material is when one spontaneously break the $\mathbb{Z}_2$ time-reversal symmetry. When this happens, one end up with system with  
domain wall separating the two sectors with $\theta = \pi$ and $\theta = -\pi$. The differences between two sectors is $j^i[\theta=\pi]-j^i[\theta=-\pi]= (e^2/h) \epsilon^{ij}E_j$. This situation is `realised' in e.g. Haldane model \cite{Haldane:1988zza} (see also section II.B. of \cite{Hasan:2010xy}).

We find no constraint on two transport coefficients $\tilde\eta$ and $\tilde\sigma$ as tensor and vector structure multiplying them, see Eq.\eqref{eq:parityOddConstitutiveReln}, vanished in equilibrium. The first coefficient $\tilde\eta$, known as the Hall viscosity, is particularly interesting: first because of its difficulty in measuring (see e.g. \cite{Hoyos:2014pba}) and recent success in hydrodynamic regime of graphene  \cite{berdyugin2019measuring,scaffidi2017hydrodynamic}. Admittedly, our computation cannot say anything about this transport coefficient. Nevertheless, it is tempting to conclude that the parity anomaly does not contribute to the Hall viscosity. A strong indication being that effective theories which produce Hall viscosity requires additional topological term of different nature \cite{Read:2010epa,Hoyos:2011ez,Gromov:2014gta} known as the Wen-Zee term (see also \cite{Golkar:2014wwa} for its relativistic generalisation). It would be interesting to understand whether or not the Wen-Zee term is a result of discrete anomaly of different kind. 

Last but not least, we hope that this paper demonstrate that, despite not affecting the Ward identity, the discrete anomaly can affect hydrodynamic transport in a nontrivial way. We only consider only one, arguably the simplest, of such examples but there are plethora of SPT phases out there. Given a rich physics that came out of the study of perturbative anomaly induced transport, this would be a very interesting direction to explore.   

\section*{Acknowledgements}

I would like to thanks Nabil Iqbal and Tin Sulejmanpasic for collaborating on related topics and various discussions. I am also very grateful to I\~naki Garc\'ia-Etxerbarria, Nakarin Lohitsiri for illuminating discussions as well as Paolo Glorioso, Nabil Iqbal and Kristan Jensen for commenting on the manuscript. I would like to thank Chulalongkorn University for hospitality during the pandemic refuge period and my work is supported by STFC grant number ST/T000708/1.

\begin{appendix}
\section{A very short interlude on APS index theorem and anomaly inflow}
\label{app:APSindex}
For self-containment sake, we will briefly outlined reasons why free massless pseudoreal fermion is anomalous as well as how APS index theorem provide a derivation of anomaly inflow mechanism for parity anomaly. All of these information are explained in more details in \cite{AlvarezGaume:1984nf,Witten:2015aba}. More examples on how APS index theorem provide anomaly inflow mechanism and computation's subtleties can be found in \cite{Witten:2019bou} and for a more familiar chiral anomaly in \cite{Kobayashi:2021jbn}. This will also help justify the reason for $\theta = \pi$ beyond the argument of \cite{Qi:2008ew}.

Start by considering a prototype of parity anomaly: the (massless) Dirac fermion in $2+1$d whose eigenvalue obtained via
\begin{equation}
  i \slashed{D} \psi_k = \lambda_k \psi_k 
\end{equation}
From this, we can obtain the partition function of free massless Dirac fermion 
\begin{equation}
  Z_\psi = \det(i\slashed{D}) = \prod_k \lambda_k
\end{equation}
which, just like any QFT partition function, is infinite. To make sense of it, we add a regulator by adding
\begin{equation}
  S_{reg} = \int d^3x \chi\left( i\slashed{D} +iM \right)\chi 
\end{equation}
for $\chi$ be a 2-components scalars that satisfies $(i\slashed{D}+iM)\chi = 0$. The resulting regulated action is 
\begin{equation}
  Z_{reg} = \prod_k \frac{\lambda_k}{\lambda_k + iM} = \exp\left( \sum_k \log \left(\frac{ \lambda_k}{\lambda_k + iM}  \right) \right)
\end{equation}
Even when $M\to +\infty$, this theory regulated theory turns out to not be invariant under time-reversal as the partition function is not real. To check the possible imaginary part, we can consider the imaginary part of the exponent 
\begin{equation}
\begin{aligned}
  \text{Im } \sum_k \log\left( \frac{\lambda_k}{\lambda_k + iM} \right) &= \sum_k \left[ -\log(\lambda_k +iM) + \log (\lambda_k -iM)  \right]\, , \\
  &= \sum_k \tan^{-1}\left( \frac{M}{\lambda_k} \right)\, , \\
  &=\frac{\pi}{2} \left[ \sum_{\lambda_k> 0} 1-\sum_{\lambda_k< 0} 1  \right] \qquad \text{as }M\to +\infty
\end{aligned}
\end{equation}
The last term in the $[..]$ is the APS $\boldsymbol{\eta}-$invariant. Computing $\boldsymbol{\eta}-$invariant requires another kind of regulator but we will not talk about that for now. What is interesting is that we can then write the regulated partition function as 
\begin{equation}
  Z_{reg} = |Z_\psi| \exp\left( i\pi \boldsymbol{\eta}/2 \right)
\end{equation}
where $|Z_\psi|$ is the modulus of the (regulated)partition function. We can see from there that the time reversal, $TiT^{-1} = -i$, transformed the regulated partition function as 
\begin{equation}
  T: \qquad Z_{reg}\to Z_{reg} \exp(-i\pi\boldsymbol{\eta})
\end{equation}
The APS index theorem state that for $\boldsymbol{\eta}-$invariant on $\CM = \d \CN$, we have 
\begin{equation}\label{eq:APSindextheorem}
  -\frac{\boldsymbol{\eta}}{2} - \int_\CN (\CP-\hat A(R))  = \mathfrak{J}
\end{equation}
where $\mathfrak{J}\in \mathbb{Z}$ is the Dirac index on $\CN$ computed with the APS boundary condition, see e.g. \cite{Witten:2015aba,Witten:2019bou}. The terms $\CP \sim F\wedge F$ is the anomaly polynomial in Section \ref{sec:derivation} (with proper normalisation to $\int \CP =1$ on a closed manifold) and $\hat A(R)$ is the Euler density. Taking the background metric to be flat, we find that 
\begin{equation}\label{eq:APSindexAsInflow}
\begin{aligned}
  Z_{reg} \exp\left(i\pi \int_\CN \CP\right) &= |Z_\psi| \exp\left(i\frac{\boldsymbol{\eta}\pi}{2} + i\pi\int_\CN \CP  \right)\, , \\
  &= |Z_\psi| (-1)^\mathfrak{J} 
  \end{aligned}
\end{equation}
the combined bulk+boundary partition function is therefore real and invariant under time-reversal. Notice that the topological term $i\pi \int \CP$ is  nothing but the SPT phase in \eqref{eq:bulkSPT} with $\theta = \pi$. 

\section{Euler density contribution to equilibrium partition function?}\label{app:EulerDensity}

As one may see else where or from \eqref{eq:APSindextheorem} that ${\boldsymbol{\eta}}/2$ and $\CP = F\wedge F$ does not only give integer $\mathfrak{J}$ but also the Euler density $\hat A(R)$. We have been conveniently ignored the contribution from $\hat A(R)$ so far. Here, we will justify the reason for that. In simple words, even if we include the Euler density, it will only contribute to transport coefficients at third order in the derivative expansions.

To verify this claim, we want to find an analgue of $A^\perp$, which for the geometric quantity turns out to involve the connection 1-form ${\bf \Gamma} = \Gamma^{\mu}_{\;\; \mu \lambda} dx^\lambda$ on $\CN$
\begin{equation}
  {\bf \Gamma}^\perp = \Gamma + \boldsymbol{\mu}_R u \, , \qquad (\muR)^a_{\;\; b} := T \nabla_b \left( \frac{u^a}{T} \right) 
\end{equation}
The curvature 2-form $\boldsymbol{R} = d{\bf \Gamma} + {\bf \Gamma}\wedge {\bf \Gamma} $ can also be written in a form analogous to the electric and magnetic field 
\begin{equation}
  {\bf R} = -u \wedge {\bf E}_R + {\bf B}_R\, , \qquad (E_R)^\mu_{\;\; \nu} = R^\mu_{\;\; \nu \alpha \beta} u^\beta dx^\alpha
\end{equation}
and the perpendicular field strength can be written (after imposing equilibrium conditions) as 
\begin{equation}
  {\bf R}^\perp = {\bf B}_R + 2\muR \omega
 \end{equation}
We can now play to game of deriving the transgression form for $\tr({\bf R}\wedge {\bf R})$. Again, with Chern-Weil theorem, we can write 
\begin{equation}
  \tr({\bf R}\wedge {\bf R}) - \tr({\bf R}^\perp\wedge {\bf R}^\perp) = d V_\CP^{grav}
\end{equation}
The flow from $\boldsymbol{\Gamma}^\perp$ to $\boldsymbol{\Gamma}$ can be written as 
\begin{equation}
  \boldsymbol{\Gamma}(\tau) = \Gamma + (1-\tau) \muR u\, , \qquad {\bf R}(\tau) = u\wedge(...)+ ({\bf B}_R+2(1-\tau) \muR \omega)
\end{equation}
The transgression form for this case is therefore 
\begin{equation}
\begin{aligned}
  V_\CP^{grav} &= \tr\left[\int^{\tau=1}_{\tau=0} d\tau \left\{ \d_\tau {\bf \Gamma}\wedge \frac{\d \CP[{\bf R}(\tau)]}{\d {\bf R(\tau)}} \right\} \right]\, , \\
  &=-\tr \left[ 2\muR u\wedge {\bf B}_R +2 [\muR]^2 u\wedge \omega       \right]
\end{aligned}
\end{equation}
All of these terms are third order in derivatives of hydrodynamic variables $(u^\mu,T)$ and metric. 

There is no known classification of parity-odd hydrodynamic with $U(1)$ symmetry at this order. The closest attempt in this direction is a classification  transport coefficients for relativistic hydrodynamic without $U(1)$ charge upto thrid order in the derivative expansions, which found 68 terms in total \cite{Grozdanov:2015kqa}. No inclusions of $U(1)$ symmetry has been done, let alone the parity-violation terms. So we hope, given the scope of this work, that this will justify leaving the consequences of Euler density part to more powerful future generations.

It should be mentioned that similar situation occur in the gravitational anomaly in the $2n$ dimensions e.g. take $n=2$ where $d\star j \propto \tr[{\bf R}\wedge {\bf R}]$. In a naive derivative counting, one would have thought that the gravitational, which only appears in $4^{th}$ order in the derivative expansions cannot affect the $\xi,\xi_B$ in \eqref{eq:Son-surowka-constiutive} which enters the Ward identity at $2^{nd}$ order. However, by requiring that the theory is consistent in a singular geometry e.g. a cone with the metric
\begin{equation}\label{eq:coneMetric}
  ds^2 = r^2 d\tau^2 + dr^2 + ds_\perp^2
\end{equation}
with transverse flat metric $ds_\perp^2$ and $\tau \sim \tau +  2\pi \delta $ with $\delta <1$ be a deficit angle signifying conical singularity. The presence of a cone inhibit a `derivative jump' that bring $4^{th}$ derivative down to $2^{nd}$ derivative level. By demanding that momentum of the combined bulk+boundary system toward the tip of the cone, $T^{\tau r}$, must vanishes, Ref. \cite{Jensen:2012kj,Jensen:2013rga} show that this condition add additional fixed remaining unknowns in \cite{Son2009,Neiman:2010zi} in terms of the mixed $U(1)$-gravitational anomaly coefficient\footnote{We emphasise that this condition is not a universal low energy relations and should only be applied to a theory that is continuously connected to free theories \cite{Glorioso:2017lcn}. The situation here is amount to say that the process of taking the geometry to be a cone in the UV and taking the RG flow to low energy limit does not commuted. }. In $2+1$ dimensions, on the other hand, one can check that putting the theory on a cone does not produce any extra constraints on transport. To see it explicitly, consider perturbing the metric of a cone in the form of
\begin{equation}
  \delta(ds^2) = h_{\tau r}(r,z) d\tau dr \,.
\end{equation}
One can check that there is no term in the ``inflowed'' action $\int_\CM V_\CP^{grav}$ contains no term linearly in $h_{\tau r}$ except the total derivative terms. Thus the condition $T^{\tau r} = 0$ of \cite{Jensen:2012kj} is trivially satisfied in $2+1$ dimension and does not producing any nontrivial constraints to the hydrodynamic transport.

That being said, while the anomaly inflow of Euler density from $\CN$ to $\CM = \d \CN$ does not constrain any transport on $\CM$ (where our fluid lives), it can have non-trivial effect if $\CM$ has a boundary. This can be understood as the Euler density on $\CN$ can be written as a gravitational term $S_{CS}[\Gamma] \sim \tr\int_\CM ({\bf \Gamma}\wedge d{\bf \Gamma} + \frac{2}{3}{\bf \Gamma}\wedge {\bf \Gamma}\wedge {\bf \Gamma})$ whose gravitational Chern-Simons coupling is fixed by $\theta =\pm \pi$ via \eqref{eq:APSindextheorem}-\eqref{eq:APSindexAsInflow}. While having no effect on the hydrodynamic modes on $\CM$, $S_{CS}[\Gamma]$ can induce anomaly inflow mechanism to $\d\CM$ \cite{Stone:2012ud}. The method of \cite{Jensen:2012kj} can then be used to determine the transport coefficients in terms of the Chern-Simons coupling, resulting in the quantised transport coefficient responsible for thermal hall conductivity \cite{Kane:1997fda}. 
\end{appendix}
 \bibliographystyle{utphys}
\bibliography{/home/nickpoovuttikul/Documents/biblio} 
\end{document}